\documentclass[twocolumn,showpacs,amsmath,amssymb,superscriptaddress]{revtex4}

\usepackage{amsmath}
\usepackage{comment}
\usepackage{graphicx, float}
\usepackage{dcolumn}
\usepackage{bm}
\usepackage{epstopdf}
\usepackage{doi,hyperref,url}

\begin{document}

\title{Dispersion of gravitational waves in cold spherical interstellar medium}

\author{D\'{a}niel Barta}
 \email{barta.daniel@wigner.mta.hu}
 \affiliation{%
    Institute for Particle and Nuclear Physics, Wigner Research Centre for Physics, Hungarian Academy of Sciences,
    Konkoly-Thege Mikl\'{o}s \'{u}t 29-33., H-1121 Budapest, Hungary
  }%
\author{M\'{a}ty\'{a}s Vas\'{u}th}
\email{vasuth.matyas@wigner.mta.hu}
\affiliation{%
	Institute for Particle and Nuclear Physics, Wigner Research Centre for Physics, Hungarian Academy of Sciences,
	Konkoly-Thege Mikl\'{o}s \'{u}t 29-33., H-1121 Budapest, Hungary
}

\begin{abstract}
We investigate the propagation of locally plane, small-amplitude, monochromatic gravitational waves through cold compressible 
interstellar gas in order to provide a more accurate picture of expected waveforms for direct detection. The quasi-isothermal gas is 
concentrated in a spherical symmetric cloud held together by self-gravitation. Gravitational waves can be treated as linearized perturbations on 
the background inner Schwarzschild spacetime. The perturbed quantities lead to the field equations governing the gas dynamics and describe the 
interaction of gravitational waves with matter. The resulted field equations decouple asymptotically for slowly varying short waves to a 
set of three PDEs of different orders of magnitude. A second-order WKB method provides transport equations for the wave amplitudes. 
The influence of background curvature already appears in the first-order amplitudes, which gives rise to diffraction. 
We have shown that the transport equation of these amplitudes provides numerical solutions for the frequency-alteration. 
The energy dissipating process is responsible for decreasing frequency. The decrease is significantly smaller than the magnitude of 
the original frequency and exhibits a power-law relationship between original and decreased frequencies.
The frequency deviation was examined particularly for the transient signal GW150914. Considering AGNs as larger 
background structures and high-frequency signals emitted by BNS mergers, the frequency-deviation grows large enough 
to be relevant in future GW-observations with increased sensitivity.
\end{abstract}

\maketitle

\section{\label{sec:intro}Introduction}
\subsection{\label{sec:motiv}Motivation and relation to other works}

Following the paper \cite{barta}, this present work is the second part of a comprehensive study that aims at investigating the dispersion of gravitational waves in 
interstellar medium. It is dedicated to explore the dumping effect of interstellar clouds (ICs) on gravitational waves (GWs), and 
to provide a more accurate picture of expected waveforms for direct detection. Most papers -- referring to the weak interaction --  
neglect the dispersive character of GWs in a medium as most of the estimates for detection are made under the assumption of GWs 
following null geodesics even in the presence of matter. The interaction is weak indeed, moreover even the densest nebul\ae{} are 
extremely thin. Nonetheless, the currently detectable GWs are expected to be of extragalactic origin, their sources are likely 
to be obscured by dust or gas in addition to the ICs passed through on their way to our ground-based interferometers. 
The subject has been actualized by the unique sensitivity of the recently upgraded Advanced LIGO and Advanced Virgo interferometric observatories, 
and the prospect to space instruments such as the proposed eLISA and DECIGO space antennas and the Einstein Telescope. Following the announcement 
of the observation of the event dubbed 'GW150914', the Era of First Discoveries has come \cite{LIGO1}. We witness a progressively dedicated search for new 
events from various astrophysical sources where the interaction may prove to be relevant.
 
The existence of GWs was predicted in 1915 on the basis of theory of general relativity. In 1974 the discovery of PSR B1913+16 by 
Hulse \& Taylor provided an unquestionably accurate, yet circumstantial evidence. For half a century a thorough examination of 
their dispersive nature seemed to be inconvenient and pointless for the strong reasons above. In their pioneering 1966 papers, Hawking and Weinberg 
\cite{hawking, weinberg} investigated the rate at which GWs are damped by a dissipative fluid in 
the case of a Robertson-Walker background spacetime. They found that the amplitude of a high-frequency gravitational wave is damped 
in a characteristic time $\eta^{-1}$, where $\eta$ is the fluid's shear viscosity. Also Isaacson \cite{isaacson} dealt 
with the high-frequency limit of GWs is of great importance and motivated numerous researches later on.

In the 1970s several authors (see \cite{weinberg, polnarev, madore1, madore2, madore3, chesters, ignatyev, anile, carter, anile2, 
gayer, sacchetti}) were engaged in studying the propagation of GWs through matter under various simplifying assumptions with 
different methods of approximations. Their primary concern was not the possible damping influence on the amplitude, but the 
modification of dispersion relation. The most commonly used two distinct models were a medium composed of deformable molecules 
with internal structure giving rise to anisotropic pressures or free particles with rare collisions described by kinetic theory. 

Polnarev \cite{polnarev}, then Chesters \cite{chesters} used methods of plasma kinetic theory for a GW in hot collisionless gas and 
derived the complex index of refraction for GW. It has been shown that in a collisionless gas the damping of GWs (that is analogous 
to collisionless Landau damping) is absent, but damping of the short-wave adiabatic perturbations is possible. The decrements of collisional 
and collisionless damping in a relativistic gas and the group velocity of a transverse GW were calculated in \cite{ignatyev} by 
Ignatyev. Madore's paper \cite{madore2} on gravitational radiation absorbed by a dissipative fluid confirmed Hawking's result in a 
more general context using eikonal approximation. Besides, he derived a dispersion relation analogous to that satisfied by plane 
electromagnetic waves in a dilute plasma \cite{madore1}, and another dispersion relation for gravitational radiation emitted from 
an almost spherical object \cite{madore3}. Both studies indicate the possible absortion by a viscous medium.

To describe the damping in kinetic theory, the rate of particle collisions has to be addressed giving rise to the 
imaginary part of the complex index of refraction. The refractive index turned out to be related to the viscosity response function in fluids 
and crystals. \cite{widom} Anile \& Breuer provided a formalism for a more accurate description of the transfer equation \cite{anile} 
for polarized gravitational radiation in terms of the Stokes parameters. Anile \& Pirronello obtained the transport equation for 
the amplitude of a GW in a dispersive medium using high-frequency approximation \cite{anile2}. Additionally, the design of Weber 
bars and subsequent resonant-mass detectors was an engineering problem of considerable importance since they were thought to be 
sensitive enough to measure the metric deviations. This problem inspired many to focus on GW refraction in condensed matters. The behaviour of 
an elastic medium under the influence of GWs was derived firstly in a gauge-independent way in terms of relative strains, and 
secondly in terms of displacements \cite{carter}. However, some authors like Gayer \& Kennel in the case of Landau damping, reached 
contradictory conclusions, particularly with respect to the dispersion relation \cite{gayer}. In the case of a parabolic Friedmann 
background, Sacchetti \& Trevese showed that the presence of matter does not affect the GW propagation at low temperature in the 
geometrical-optics limit $\mathcal{O}(\eta)$, however a second-order Wentzel--Kramers--Brillouin (WKB) -- so-called 
''post-geometrical optics'' -- approximation revealed a plasmalike dispersion relation in the $\mathcal{O}(\eta^{2})$, 
where $\eta$ was a small parameter \cite{sacchetti}. Recently, a study \cite{keresztes} of perfect-fluid perturbations of Kantowski--Sachs models with a positive cosmological constant has concluded that in contrast with the Friedmann case, one of the two gravitational degrees of freedom is coupled to the matter density perturbations, and decouples only in the geometrical optics limit. There, the dynamics is encompassed in six evolution equations, representing forced oscillations and two uncoupled damped oscillator equations.

In the following two decades interest in the field slightly diminished and only a few new articles were published 
(see \cite{widom, efroimsky, dolan, ehlers, ehlers2, prasanna}). Efroimsky's intention in \cite{efroimsky} to show on an arbitrary curved 
background, in an arbitrary media, that a natural low-frequency cut-off depends only on the peculiarities of the observer is 
noteworthy. Dolan examined the scattering of a low-frequency GW by a massive compact body in vacuum and found that the partial-wave 
cross section agrees with the cross section derived via perturbation-theory methods. \cite{dolan}

Undoubtedly the most remarkable of these papers are a series of studies \cite{ehlers, ehlers2, prasanna} published by Ehlers, Prasanna and Breuer 
either in collaboration or on their own. The first of these works which in some ways reviewed the results of \cite{anile}, revealed two degenerate 
modes of polarization -- one represented GWs, whereas another one described non-propagating density and vorticity perturbations -- by the dispersion 
relation for small-wavelength, small-amplitude GWs propagating through an arbitrary background dust spacetime. The following paper \cite{ehlers2} 
generalized the background to perfect fluids, through which one more mode for sound waves was identified; all modes, but the doubly degenerate 
zero-frequency matter mode exhibited propagation along the null geodesics of the background. The work was further extended by Prasanna \cite{prasanna} 
to include dissipative terms of shear and bulk viscosity in the energy momentum tensor. In order to avoid loss of generality, assumption on symmetry of 
background had never been initiated. Generality however also entails several disadvantages: due to the lack of specifically detailed background, alterations 
in the wave's amplitude and frequency cannot be determined. The procedure developed in \cite{ehlers, ehlers2, prasanna} 
was generalized to curved backgrounds and is largely similar to that of Sv\'{\i}tek's approach to the damping of GWs in dust cloud \cite{svitek}. 
There the geodesic deviation equation stemed from the periodic oscillations produced within the molecules by the incoming waves. The oscillations 
themselves produce GWs that were composed with the original waves propagating through the cloud from a distant source.

\subsection{\label{sec:Overview}Overview of content}

This present work is largely based on these three outstanding studies (\cite{ehlers, ehlers2, prasanna}) and is a sequel to our preceding paper \cite{barta} where 
an appropriate background metric has been set up into which metric perturbations could be embedded. There a self-gravitating spherical shell of collisionless gas 
composed by cold neutral hydrogen molecules in thermal equilibrium was modeled in general relativistic framework. This framework permits one to make up for the aforementioned disadvantages. 

Sec. \ref{sec:background} discusses the background and its perturbations in detail where the physical properties of the molecular cloud were taken from \cite{murray}. In Sec. \ref{sec:fieldeq}, the field equation is defined wherein the perturbed quantities obtained in Sec. \ref{sec:linearpert} give rise to the field equations governing the gas dynamics and express the interaction of GWs with matter. In this paper we do not take into account the back reaction of perturbations on the background spacetime but consider the background given in Sec. \ref{sec:background}. It has been shown that the derived equations in the dust-filled flat-spacetime case correspond to that of \cite{ehlers}.

Sec. \ref{sec:WKB-optics} gives rise to a subject for discussion for Wentzel--Kramers--Brillouin analysis and expansions of geometric optics. In this context, in Sec. \ref{sec:WKB}, the field equations decoupled asymptotically for monochromatic high-frequency waves to a set of partial differential equations (PDEs) of different orders of magnitude by developing a second-order (i.e. ’post-geometrical optics’) WKB method. Distinction between regular and singular modes was made and the amplitudes of the perturbation were separated into primary and secondary parts. In the regular case it was found that the primary amplitudes follow null characteristics whereas the transport of secondary amplitudes depends upon the density in geometric optics limit. The explicit form of the arisen differential operators $\mathcal{L}$ is given in Sec. \ref{sec:diffops}. We has demonstrated that GWs within geometrical-optics approximation follow null geodesics, even in the presence of medium, while their dispersive character is exhibited, and has been adequately studied in the post-geometrical optics limit. In the lowest order, the crossing waves propagate as in vacuum. The influence of background curvature occurs only in higher orders, which exhibit diffraction. In Sec. \ref{sec:transport-eqs} the transport equation for the first-order primary amplitudes are formally given. 

In Sec. \ref{sec:num-solution} the transport equation is converted into a PDE which is numerically solved where the first-order primary amplitudes are assumed to change in a sinusoidal manner over retarded time. However extremely small in value, the frequency shows a decreasing behaviour that can be traced back to the nature of dissipative interaction of the GW with the surrounding matter. From the decomposition of any given signal into varying sinusoidal components by Fourier analysis, we can construct the changes in frequency of all the sinusoids for all the frequencies. In Sec. \ref{sec:gw-analysis} we summon the match-filtering technique to our aid. \cite{apostolatos} This well-established concept in GW data analysis correlates a known template with an unknown signal to detect the presence of the template in the unknown signal. For its importance, the reconstructed time series of the transient GW signal 'GW150914' as a known template is taken into examination.\cite{LIGO1,LIGO2} Our first and foremost priority is to measure the deviation of this signal by an overlap function.  Secondly, is to bring to light in what types of possible GW-sources the effect of interaction is expected to be powerful enough to be taken into account for future examination.

\subsection{\label{sec:notations}Notations and conventions}
We shall use geometrical units, i.e. the speed of light $c$ in vacuum and the gravitational constant $G$ are set to unity by an appropriate 
choice of units so that the $\kappa \equiv 8\pi G/c^3 = 8\pi$. Spatial coordinates are labeled with Latin indices $a,\, b \ldots$, whereas 
the ones belonging to 4-vectors and tensors are labeled with Greek indices $\alpha,\, \beta \ldots$. We conventionally denote time 
coordinate by the index 0, while spatial coordinates are denoted by indices running from 1 to 3.  According to Einstein's notational convention, 
we are summing over all of the possible values of that index variable which occurs twice in a single term, once in an upper and once in a 
lower position. Lower indices stand for covariant quantities, upper ones for contravariant ones, whereas both in a single term denote mixed 
variance. Symmetrization of indices is indicated by $(\,)$, anti-symmetrization by $[\,]$. Partial derivatives are indicated by $\partial_{\mu}$ or a 
comma, single and repeated covariant derivatives by $\nabla_{\mu},\, \nabla_{\mu\nu}$ or by a semicolon. The metric signature is chosen as $(-\, +\, +\, +)$ 
according to the Pauli space-like convention.

\section{\label{sec:background}Background and perturbations}
\subsection{\label{sec:fieldeq}Field equation and dynamics of gas}
Let some smooth, non-degenerate, symmetrical metric tensor $g_{\mu\nu}$ be given on some differentiable mainfold $\mathcal{M}$ 
which existed before the train of GWs came, be disturbed by the small linear perturbation $h_{\mu\nu}$; then another 'total' metric
\begin{equation} \label{total-metric-eq}
\tilde{g}_{\mu\nu} \equiv g_{\mu\nu} + h_{\mu\nu}
\end{equation}
is called into being. For the covariant components, one has to set 
\begin{equation} \label{inverse-total-metric-eq}
	\tilde{g}^{\mu\nu} \equiv g^{\mu\nu} - h^{\mu\nu} + \mathcal{O}(h^{2})
\end{equation}
in order to obey the necessary condition $\tilde{g}_{\mu\rho}\tilde{g}^{\rho\nu} = \delta_{\mu}^{\nu}$ for the new metric. It is requisite to stress that the tensor field $h^{\mu\nu}$ 
is small in the sense that the term denoted as $\mathcal{O}(h^{2})$ is small against the preceding term. Let us consider the spacetime $(\mathcal{M},\, g_{\mu\nu})$ of 
a remote and isolated IC filled with cold perfect fluid, obeying the field equation
\begin{equation} \label{field-eq}
R_{\mu\nu} = \kappa(\rho + p)\left(u_{\mu}u_{\nu} + \frac{1}{2}g_{\mu\nu}\right) - \kappa p g_{\mu\nu}
\end{equation}
where $\rho,\, p > 0$ are the density and pressure respectively; $u^{\mu}$ is the normalized 4-velocity of a gas element which in turn satisfies 
the geodesic equation
\begin{equation} \label{geodesic-eq}
u^{\mu}\nabla_{\mu}u^{\nu} = 0
\end{equation}
along with the continuity equation
\begin{equation} \label{continuity-eq}
u^{\mu}\nabla_{\mu}(\rho + p)u^{\nu} + g^{\mu\nu}\nabla_{\mu}p = 0.
\end{equation}
Assume that the cloud is in thermal equilibrium and forms a sphere under the influence of its own gravitation. The total metric $\tilde{g}_{\mu\nu}$, therefore refers to the distorted pre-metric $g_{\mu\nu}$ of an isolated Schwarzschild spacetime exposed in external fields of GWs. The line element associated with such a background was given as \cite{barta}
\begin{equation} \label{line-element}
\begin{array}{l}
\mathrm{d}s^{2} = \displaystyle -\frac{c_{\text{s}}^{2}}{4}\left(1 + \frac{c_{\text{s}}^{2}}{4}\frac{r^{2}}{R^{2}}\right)\mathrm{d}t^{2} + 
\exp\left(-\frac{c_{\text{s}}^{2}}{2}\frac{r^{2}}{R^{2}}\right)\mathrm{d}r^{2} \\[10pt]
\hspace{28pt} +\ \displaystyle r^{2}(\mathrm{d}\vartheta^{2} + \sin^{2}\vartheta \mathrm{d}\varphi^{2})
\end{array}
\end{equation}
for polytropic equation of state (EoS) and corresponding profile of density
\begin{equation} \label{EoS}
p = c_{\text{s}}^{2}\rho^{\gamma}, \hspace{3pt} \rho(r) = \frac{3c_{\text{s}}^{2}}{16\pi}\left(\frac{16r^2}{R^2}-1\right)\left(\frac{8r^2}{3R^2}-1\right) 
\end{equation}
wherein the polytropic exponent $\gamma$ is unitary and $c_{\text{s}}$ is the isothermal speed of sound within the medium. The geodesic equation 
(\ref{geodesic-eq}) along with the plausible assumption that radial motion is irrelevant from our perspective imply that massive particles move along circular 
orbits to that end, the eccentric contributions to the time-averaged velocity cancel each other out. The components of 4-velocity of a particle in the equatorial 
plain $\vartheta = \pi/2$ are thus given by
\begin{equation}
\left[u^{\mu}\right] = \left[1,\, 0,\, 0,\, \frac{c_{\text{s}}^{2}}{4}\frac{r}{R}\sin\vartheta\right]
\end{equation}
in the coordinate system $(t, r, \vartheta, \varphi)$. The continuity equation (\ref{continuity-eq}) requires that 
\begin{equation} \label{c*-eq}
u_{\mu}u^{\mu} = -\frac{c_{\text{s}}^{2}}{4} + \mathcal{O}(c_{\text{s}}^{4}/c^2) \equiv -c_{*}^{2}
\end{equation}
where it is sufficient to keep only the leading order term. Due to symmetry, the connection has only four non-zero components
\begin{equation}
\begin{array}{ll}
\Gamma^{0}_{01} = \displaystyle \left(1 + \frac{c_{s}^{2}}{4c^{2}}\frac{r^{2}}{R^{2}}\right)^{-1}\frac{c_{s}^{2}}{4c^{2}}\frac{r}{R^{2}}, & \quad \Gamma^{1}_{11} = \displaystyle -\frac{c_{s}^{2}}{2c^{2}}\frac{r}{R^{2}} \\[10pt]
\Gamma^{1}_{00} = \displaystyle \frac{c_{s}^{4}}{16c^{4}}\frac{r}{R^{2}}\exp\left(\frac{c_{s}^{2}}{2c^{2}}\frac{r^{2}}{R^{2}}\right), & \quad \Gamma^{2}_{12} = \displaystyle \frac{1}{r}.
\end{array}
\end{equation}

\subsection{\label{sec:linearpert}Linear perturbations of gas}
Let a 'background field' be denoted as ($g_{\mu\nu},\, u^{\mu},\, \rho,\, p$) and with overhead hat a linearized perturbation as ($\hat{g}_{\mu\nu},\, \hat{u}^{\mu},\, \hat{\rho},\, \hat{p}$). According to linearization stability, ($g_{\mu\nu} + \delta\hat{g}_{\mu\nu},\, u^{\mu} + \delta\hat{u}^{\mu},\, \rho + \delta\hat{\rho},\, p + \delta\hat{p}$) will approximate a solution of the field equation, at least in a compact part of spacetime, provided the constant numerical factor $\delta \ll 1$ is sufficiently small. So far no straightforward criterion has been known to properly define when $\delta$ is 'sufficiently' small. \cite{ehlers} Nevertheless, by estimating the order of magnitude of several terms in manner of Isaacson \cite{isaacson} one can conclude that $\delta$ has to be taken to be higher order than $\varepsilon^{2}$ in the lowest WKB order, seeing that for larger amplitudes $\delta \geq \varepsilon^{3}$, the error due to the restriction to linearized perturbations is expected to be larger than the leading WKB amplitudes.

Suppose that ($\hat{g}_{\mu\nu},\, \hat{\rho},\, \hat{p}, \hat{u}^{\mu}$) is a solution of the linearized equations corresponding to (\ref{field-eq}--\ref{geodesic-eq}) at a background solution ($g_{\mu\nu},\, u^{\mu},\, \rho,\, p$). Let $\varepsilon$ denote a small ratio between the scale of 
variation of the perturbed variables to that of the background. Let us associate the potential $h_{\mu\nu}$ of the radiation in (\ref{total-metric-eq}) with a 
small linear perturbation $\varepsilon \hat{g}_{\mu\nu}$ which allows us to express all the quantities uniformly as
\begin{equation} \label{perturbed-quantities-eq}
\begin{array}{ll}
& \tilde{g}_{\mu\nu} = g_{\mu\nu} + \varepsilon \hat{g}_{\mu\nu} \\
& \tilde{u}_{\mu} = u_{\mu} + \varepsilon \hat{u}_{\mu} \\
& \tilde{\rho} = \rho + \varepsilon \hat{\rho} \\
& \tilde{p} = p + \varepsilon \hat{p}.
\end{array}
\end{equation}
In the same manner, the covariant components of the total energy-momentum tensor is the sum 
\begin{equation}
\tilde{T}_{\mu\nu} = T_{\mu\nu} + \hat{T}_{\mu\nu}
\end{equation}
where the un-perturbed part is the stress--energy tensor of perfect fluids, derived from the field equation (\ref{field-eq}), whereas the other one is being the contribution of the radiation, called as Isaacson stress-energy tensor
\begin{equation}
\hat{T}_{\mu\nu} = \frac{1}{32 \pi}\langle h_{\alpha\beta,\mu} h_{\alpha\beta,\nu} \rangle.
\end{equation}
An interesting situation occurs when no further sources of energy besides gravitational waves are present. The tensor field $\hat{g}_{\mu\nu}$ consists of modes with wavelengths not exceeding some maximal scale $L$. The background field varies on this characteristic length $L$ that is related to the energy density of waves; in other words $L^{2} \approx (\lambda/\varepsilon)^{2}$ where $\lambda$ is the scale, 'typical' wavelength on which the 'typical' amplitude $\varepsilon$ of the wave $h_{\mu\nu}$ varies relatively slowly, compared to the scale on which the background field varies. In this simple situation but a single small parameter $\varepsilon = \lambda/L$ is sufficient to develop the field equation. \cite{isaacson} Contrastingly, in case the background is more curved than that produced by the wave itself, besides $\varepsilon$ another small independent parameter $\eta = \lambda/L$ is required. Suppose that $|h_{\mu\nu,\lambda}| = L^{-1}|h_{\mu\nu}|$. Be that as it may, in this later case $L^{2} < (\lambda/\varepsilon)^{2}$, expressly, $\eta < \varepsilon$. Developing to the first and to the second order, one obtains, respectively, the geometrical optics and the self-interaction phenomena in both situations with the wave travelling along null geodesics. \cite{madore1}

Based on eq. (\ref{inverse-total-metric-eq}), the appropriate contravariant tensors are related to the covariant ones via the identity $\tilde{g}^{\mu\alpha}\tilde{g}_{\alpha\nu} = (g^{\mu\alpha} - \varepsilon \hat{g}^{\mu\alpha})(g_{\alpha\nu} + \varepsilon \hat{g}_{\alpha\nu}) = \delta^{\mu}_{\nu} + \mathcal{O}(\varepsilon^{2})$, 
hence the inverse total metric is ought to be expressed as
\begin{equation}
\tilde{g}^{\mu\nu} = g^{\mu\nu} - \varepsilon \hat{g}^{\mu\nu}.
\end{equation}
Consequently, the contravariant notions of perturbed quantities of different tensorial order are
\begin{equation}
\begin{array}{ll}
& \hat{g}^{\alpha\beta} = g^{\alpha\rho}g^{\beta\sigma}\hat{g}_{\rho\sigma} \\
& \hat{u}^{\alpha} = g^{\alpha\beta}\hat{u}_{\beta} + \hat{g}^{\alpha\beta}u_{\beta}. \\
\end{array}
\end{equation}
Let $\nabla$ be the covariant derivative associated with some torsionless and metric-compatible connection $\Gamma^{\kappa}_{\mu\nu}$, then the perturbation of this connection is a tensor $\hat{\Gamma}^{\kappa}_{\mu\nu}$ given \cite{ehlers} by
\begin{equation}
\hat{\Gamma}^{\gamma}_{\alpha\beta} = \frac{1}{2}g^{\gamma\delta}(\nabla_{\beta}\hat{g}_{\delta\alpha} + \nabla_{\alpha}\hat{g}_{\delta\beta} - \nabla_{\delta}\hat{g}_{\alpha\beta}).
\end{equation}
Based on this scheme of work, the Ricci identity imply
\begin{equation} \label{perturbed-ricci-eq1}
\mkern-\thinmuskip \hat{R}_{\alpha\beta} = \frac{1}{2}\left(\nabla^{\gamma}_{\alpha}\hat{g}_{\beta \delta} + \nabla^{\delta}_{\beta}\hat{g}_{\alpha \delta} - \nabla^{\delta}_{\delta}\hat{g}_{\alpha\beta} - \nabla_{\alpha\beta}\hat{g}_{\delta\gamma}g^{\delta\gamma}\right).\!\!
\end{equation}
Likewise, by retaining only terms of the first order in the linear approximations of eqs. (\ref{field-eq}) and (\ref{c*-eq}), one obtains
\begin{equation} \label{perturbed-ricci-eq2}
\begin{array}{ll}
\hat{R}_{\alpha\beta} = & \kappa(\hat{\rho}+\hat{p})\left(u_{\alpha}u_{\beta} + \frac{1}{2}\mathring{g}_{\alpha\beta}\right) - \kappa\hat{p}g_{\alpha\beta} \\[3pt]
& +\ \kappa(\rho+p)\left(2u_{(\alpha}\hat{u}_{\beta)} + \frac{1}{2}\hat{g}_{\alpha\beta}\right) \\
\end{array}
\end{equation}
and
\begin{equation}
2u^{\alpha}\hat{u}_{\alpha} = \hat{g}_{\alpha\beta}u^{\alpha}u^{\beta}
\end{equation}
respectively. From these last two equations the perturbation of the remaining quantities can be expressed as
\begin{equation} \label{perturbed-density-eq}
\hat{\rho} + \hat{p} = -\left(2\kappa^{-1}\hat{R}_{\alpha\beta} + (\rho + p)\hat{g}_{\alpha\beta}\right)c_{*}^{-2}u^{\alpha}u^{\beta}
\end{equation}
and
\begin{equation} \label{perturbed-velocity-eq}
\hat{u}_{\alpha}  = \frac{1}{2}\left(\hat{g}_{\alpha\gamma} -\frac{2\hat{R}_{\beta\gamma} + \kappa(\rho - p)\hat{g}_{\beta\gamma}}{\kappa c_{*}^{4}(\rho + p)}\mathcal{P}^{\beta}_{\alpha}\right)u^{\gamma}.
\end{equation}
Up to this point, our results entirely corresponded to that of Ehlers and Prasanna \cite{ehlers, ehlers2}. Now, however, we shall define not one, but a pair of appropriate tensors
\begin{equation} \label{projectors-eq}
\mathcal{P}^{\beta}_{\alpha} \equiv c_{*}^{2}\delta^{\beta}_{\alpha} + u_{\alpha}u^{\beta}, \quad \mathcal{Q}^{\beta}_{\alpha} \equiv c_{*}^{2}\delta^{\beta}_{\alpha} - u_{\alpha}u^{\beta}
\end{equation}
which significantly simplify expressions to appear later on. Adopted from \cite{ehlers}, $\mathcal{P}$ projects onto the subspaces of the tangent spaces that are orthogonal to $u_{\alpha}$ whilst to the auxiliary tensor $\mathcal{Q}$ may not carry such evident geometrical meaning. The set of eqs. (\ref{perturbed-ricci-eq2}--\ref{perturbed-velocity-eq}) yields the key equation that drives the spatio-temporal evolution of linearized perturbations. Making use of the preceding notation \eqref{projectors-eq}, it can be appreciably simplified to
\begin{equation} \label{perturbation-governing-eq}
\begin{array}{l}
\!\! \displaystyle \left[\mathcal{P}^{\gamma}_{\alpha}\mathcal{P}^{\delta}_{\beta} - \mathcal{Q}_{\alpha\beta}u^{\gamma}u^{\delta}\right]\left[\hat{R}_{\gamma\delta} + \frac{1}{2}\kappa(\rho - p)\hat{g}_{\gamma\delta}\right] = \\[3pt]
\! \kappa c_{*}^{4}(c^{2}\rho - 3p)\hat{g}_{\alpha\beta} + 2\kappa(c_{*}^{2}pu^{\gamma}\hat{u}_{\gamma} - \hat{p})g_{\alpha\beta}.
\end{array}
\end{equation}
It is important to highlight two noteworthy details about the equation:
\begin{enumerate}
	\item  In flat spacetime $c_{*}$ had been equal to $1$, thus the last term in the right-hand side would perish. In our case $c_{*} \ll 1$, subsequently  the term $\mathcal{O}(c_{*}^{4})$ might be disregard as well.
	\item If one had assumed the IC to be made of nothing, but dust, such terms would vanish just as well where the pressure or its perturbation appear.
\end{enumerate}
In consequence, our result in the flat dust-filled spacetime limit would be identical to the formula (21) in \cite{ehlers}. By its very nature, eq. \eqref{perturbation-governing-eq} inherently satisfies the conditions (\ref{geodesic-eq}--\ref{continuity-eq}). However, in deriving \eqref{perturbation-governing-eq}, no restriction on the perturbations has been imposed. In order to eliminate redundant components, one can take the liberty to impose a gauge condition. Conventionally, either the de Donder (as in \cite{sacchetti}) or the Landau gauge (as in \cite{ehlers}) have been favored. The earlier (also known as harmonic coordinate condition ) ultimately requires the product $\Gamma^{\beta}_{\gamma\gamma}\hat{g}_{\alpha\beta}$ to vanish whereas the later one imposes
\begin{equation} \label{Landau-gauge-eq}
\hat{g}_{\alpha\beta}u^{\beta} = 0.
\end{equation}
This later gauge condition is particularly suitable for the fact that the metric perturbation contracts to the unperturbed 4-velocity in several terms of the perturbed field equation. By requiring \eqref{Landau-gauge-eq} to hold, \eqref{perturbation-governing-eq} reduces to the form
\begin{equation} \label{perturbation-governing-in-Landau-gauge-eq}
\begin{array}{l}
\displaystyle \left[\mathcal{P}^{\gamma}_{\alpha}\mathcal{P}^{\delta}_{\beta} - \mathcal{Q}_{\alpha\beta}u^{\gamma}u^{\delta}\right]\left[\delta^{\mu}_{(\gamma}\nabla^{\nu\phantom{\mu}}_{\delta)} - \delta^{\mu}_{\gamma}\delta^{\nu}_{\delta}\nabla^{\lambda}_{\lambda} - g^{\mu\nu}\nabla_{\gamma\delta}\right]\hat{g}_{\mu\nu} \\[5pt]
\ = \kappa c_{*}^{4}(\rho - 5p)\hat{g}_{\alpha\beta} - 4\kappa c_{*}^{4}\hat{p}g_{\alpha\beta}
\end{array} 
\end{equation}
where the operator on the left-hand side of this formula maps the space of symmetric 'spatial' tensor field into itself. This basic equation restricted by \eqref{Landau-gauge-eq} is an unconstrained system of six coupled ordinary differential equation of second order for six unknown variables. A solution is therefore specified by twelve functions of three variables. Since the restricted gauge freedom consists of four functions, the intrinsic freedom of the perturbation amounts to eight functions corresponding to four degrees of freedom.

\section{\label{sec:WKB-optics}WKB expansions of geometric optics}
\subsection{\label{sec:WKB}WKB analysis for monochromatic high-frequency field perturbations}
The set of field equations \eqref{perturbation-governing-in-Landau-gauge-eq} for metric perturbations results a system of coupled linear first-order partial differential equations (PDEs). A strategy for finding a unique closed-form analytical solution for arbitrary initial data is based on decoupling the set order by order by small parameter $\varepsilon$. Let a linear parabolic PDE
\begin{equation} \label{general-PDE}
\mathcal{D}(\textbf{x}, \nabla)\hat{g}_{\mu\nu} = (\mathcal{D}_{2}^{\alpha\beta}(\textbf{x})\nabla_{\alpha\beta} + \mathcal{D}_{1}^{\alpha}(\textbf{x})\nabla_{\alpha} + \mathcal{D}_{0}(\textbf{x}))\hat{g}_{\mu\nu} = 0
\end{equation}
in $\textbf{x} \equiv (t,\vec{x})$ be given for some function $\hat{g}_{\mu\nu}: \mathbb{R}^{n \times n} \to \mathbb{R}^{m \times m}$ where $\mathcal{D}_{2}, \mathcal{D}_{1}, \mathcal{D}_{0} \in \mathbb{R}^{m \times m}$ are matrix-valued, smooth functions with real entries which act on the six-dimensional space of metric perturbations. \cite{ehlers2} Geometrical optics emerges as a short-wavelength limit for solutions to the PDE \eqref{general-PDE}. Accordingly, let us specialize metric perturbations $\hat{g}_{\mu\nu}$ to locally plane, monochromatic, high-frequency fields. Assuming formal solutions can be locally approximated in a successive procedure by plane wave
\begin{equation} \label{WKB-eqs1}
\hat{g}_{\mu\nu}(\textbf{x},\varepsilon) = A_{\mu\nu}(\textbf{x},\varepsilon)\exp[i\varepsilon^{-1}\psi(\textbf{x},\varepsilon)],
\end{equation}
a WKB ansatz can be constructed, provided that for any $n \in \mathbb{N}$, amplitude $A_{\mu\nu}$ and phase $\varepsilon^{-1}\psi$ take the form of asymptotic series expansions
\begin{equation} \label{WKB-eqs2}
\begin{array}{lcl}
\displaystyle A_{\mu\nu}(\textbf{x},\varepsilon) \sim  \sum_{\substack{n = 0}}^{\infty}\left(\frac{\varepsilon}{i}\right)^{n}A_{\mu\nu}^{(n)}(\textbf{x}) \\
\displaystyle \psi(\textbf{x},\varepsilon) \sim  \sum_{\substack{n = 0}}^{\infty}\varepsilon^{n}\psi^{(n)}(\textbf{x})
\end{array}  
\end{equation}
in the limit $\varepsilon \to 0$. It is evidently expressed that as long as the parameter $\varepsilon$ is small, the amplitude varies slowly in comparison with the rapid oscillation of the phase. On account of the gauge-fixing condition \eqref{Landau-gauge-eq}
\begin{equation} \label{Landau-gauge-eq2}
A^{(n)}_{\mu\nu}u^{\nu} = 0
\end{equation}
is required to hold for any non-negative $n$, in other words, any $n$-order wave amplitude is transversal to the direction of propagation. The wave covector is defined to be $\varepsilon^{-1}l_{\mu}$ where
\begin{equation} \label{wavecovector}
l_{\mu} \equiv \nabla_{\mu}\psi, \quad k_{\mu} = \mathcal{P}_{\mu\nu}l^{\nu}
\end{equation}
and the angular frequency relative to the unperturbed matter flow is $-\varepsilon^{-1}u^{\mu}l_{\mu}$ which itself would be denoted by
\begin{equation} \label{omega}
\omega \equiv 2\pi f \equiv -u^{\mu}l_{\mu} = -k_{0}.
\end{equation}
One may re-arrange the expression $\mathcal{D}\left(A_{\mu\nu}\exp[i\psi/\varepsilon]\right)$ after having the ansatz (\ref{WKB-eqs1}--\ref{WKB-eqs2}) inserted  into (\ref{perturbation-governing-in-Landau-gauge-eq}) and taken into account the choice of gauge \eqref{Landau-gauge-eq2} by requiring the terms of order $1, \varepsilon, \varepsilon^{2}, \ldots$ of the resulting formal series
\begin{equation} \label{PDE-in-series}
	\left(\mathcal{L}^{(0)}{}^{\alpha\beta}_{\mu\nu} + \frac{\varepsilon}{i}\mathcal{L}^{(1)}{}^{\alpha\beta}_{\mu\nu} + \left(\frac{\varepsilon}{i}\right)^{2}\mathcal{L}^{(2)}{}^{\alpha\beta}_{\mu\nu}\right)\left(A_{\alpha\beta}^{(0)} + \frac{\varepsilon}{i}A_{\alpha\beta}^{(1)} + \ldots\right) = 0
\end{equation}
to vanish separately.
In the general case of $p$th order differential operator $\mathcal{D}$, any $\mathcal{L}^{(j)}$ is a linear differential operator of order $j$ and depends on $l_{\alpha}$ and its derivatives up to order $j$ for any positive integer $j < p - 1$. $\mathcal{L}^{(0)}, \ldots, \mathcal{L}^{(p)}$ are called the symbols of $\mathcal{D}$, roughly speaking, replacing each partial derivative by the variable $l_{\alpha}$. The highest-order terms of the symbol, the principal symbol
\begin{equation} 
\displaystyle \mathcal{L}^{(0)} = \mathcal{D}_{2}^{\alpha\beta}l_{\alpha}l_{\beta},
\end{equation}
almost completely controls the qualitative behavior of solutions of \eqref{general-PDE}. Whereas the next-to-leading order symbol ought to be
\begin{equation} 
\displaystyle \mathcal{L}^{(1)} = 2\mathcal{D}_{2}^{\alpha\beta}l_{\alpha}\nabla_{\beta} + \mathcal{D}_{2}^{\alpha\beta}\nabla_{\alpha}l_{\beta} + \mathcal{D}_{1}^{\alpha}l_{\alpha}.
\end{equation}
In fact, the principal parts of $\mathcal{L}^{(0)}, \ldots, \mathcal{L}^{(p)}$ are determined by the principal part of the original operator. Zeros of the principal symbol correspond to the characteristics of \eqref{general-PDE}: as long as $\psi$ obeys the characteristic equation
\begin{equation} \label{characteristic-eq}
	\det\mathcal{L}^{(0)} = \det\left(\mathcal{D}_{2}^{\alpha\beta}(\textbf{x})l_{\alpha}l_{\beta}\right) \equiv Q(\textbf{x},\textbf{l}) = 0,
\end{equation}
the lowest, 'zeroth' order equation
\begin{equation} \label{zeroth-order}
\mathcal{L}^{(0)}{}^{\alpha\beta}_{\mu\nu}A_{\alpha\beta}^{(0)} = 0,
\end{equation}
resulting from series \eqref{PDE-in-series}, admits non-trivial solutions $A_{\alpha\beta}^{(0)}$. $Q$ is known as the characteristic form of $\mathcal{D}$, some homogenous polynomial of degree $2m$ in the variables $l^{\alpha}$ whose coefficients depend on $x^{\alpha}$. From geometrical point of view, $Q$ is a function on phase space $\pi \equiv \{(\textbf{x},\textbf{l})\}$, wherein those characteristic set of points which satisfying \eqref{characteristic-eq} consists of several hypersurfaces which may intersect or touch each others.
The symbols corresponding to field equations \eqref{perturbation-governing-in-Landau-gauge-eq} expressed through \eqref{general-PDE} are of the forms
\begin{widetext}
	\begin{equation} \label{L-operators}
		\begin{array}{ll}
			\mathcal{L}^{(0)}{}^{\alpha\beta}_{\mu\nu} \equiv & \!\! -2{\mathcal{P}^{\alpha}}_{(\mu}l^{\beta}k_{\nu)} + \mathcal{P}^{\alpha}_{\mu}\mathcal{P}^{\beta}_{\nu}l^{2} + g^{\alpha\beta}(k_{\mu}k_{\nu} - \omega^{2}\mathcal{Q}_{\mu\nu}) \\[10pt]
			
			\mathcal{L}^{(1)}{}^{\alpha\beta}_{\mu\nu} \equiv & \!\! -4\mathcal{P}^{\gamma}_{(\mu}\mathcal{P}^{(\alpha}_{\nu)}(l^{\beta)}\nabla_{\gamma} + \nabla^{\beta)}l_{\gamma}) + 2\mathcal{P}^{(\alpha}{}_{(\mu}k{}_{\nu)}\nabla^{\beta)} 
			+ 2\mathcal{P}^{(\alpha}_{(\mu}\mathcal{P}^{\beta)}_{\nu)}(\nabla_{\textbf{l}} + \theta/2) +2\omega\mathcal{Q}_{\mu\nu}\nabla^{\beta}\mathring{u}^{\alpha} \\
			& \!\! - \mathcal{P}^{\alpha\beta}\Big[\mathcal{P}^{\gamma}_{(\mu}\mathcal{P}^{\delta}_{\nu)}\nabla_{\delta}l_{\gamma} + \mathcal{P}^{\gamma}{}_{(\mu}k{}_{\nu)}\nabla_{\gamma} - \mathcal{Q}_{\mu\nu}(2\omega\nabla_{\textbf{u}} - u^{\gamma}\nabla_{\textbf{u}}l_{\gamma})\Big] \\[10pt]
			
			\mathcal{L}^{(2)}{}^{\alpha\beta}_{\mu\nu} \equiv & \!\! \mathcal{P}^{\gamma}_{\mu}\mathcal{P}^{\delta}_{\nu}(2g^{\alpha}_{\lambda}g^{\beta}{}_{(\gamma}\nabla_{\delta)}\nabla^{\lambda} + g^{\alpha}_{\gamma}g^{\beta}_{\delta}\nabla_{\lambda}\nabla^{\lambda}) - 4\kappa c_{*}^{2}\hat{p}g^{\alpha}_{\mu}g^{\beta}_{\nu} 
			- \kappa c_{*}^{4}(\rho - 5p)g^{\alpha}_{\mu}g^{\beta}_{\nu} - g^{\alpha\beta}(\mathcal{P}^{\gamma}_{\mu}\mathcal{P}^{\delta}_{\nu} - \mathcal{Q}_{\mu\nu}u^{\gamma}u^{\delta})\nabla_{\gamma\delta},
		\end{array}
	\end{equation}
\end{widetext}
where in addition to the abbreviations $\theta \equiv \nabla^{\mu}l_{\mu}, \nabla_{\textbf{u}} 
\equiv u^{\mu}\nabla_{\mu}, \nabla_{\textbf{l}} \equiv l^{\mu}\nabla_{\mu}$, identities (\ref{wavecovector}--\ref{omega}) are applied. Clearly, the interaction between matter and GW is weak, it manifests itself only in second-order. The source term consists of two different kinds of terms: $\kappa c_{*}^{4}(\rho - 5p)g^{\alpha}_{\mu}g^{\beta}_{\nu}$ represents simply the direct influence of matter on the metric perturbations whereas the term of $\hat{p}$ refers to the contribution of pressure fluctuations which can be regarded as radiation reaction, therefore the latter has only a negligibly small influence.

\subsection{\label{sec:eikonal}Eikonal equation and rays}
There are several ways to formulate geometrical optics. One plausible formulation is where a PDE for amplitude and phase, the eikonal equation
\begin{equation} \label{eikonal-eq}
	\mathcal{H}(\textbf{x}, \textbf{l}) = 0
\end{equation}
can be encountered that describes the phase fronts of the wave. If \eqref{eikonal-eq} describes locally such a hypersurface $\Sigma$ in $\pi$, over some domain $\mathbb{R}^{n}$, on $\Sigma: \partial\mathcal{H}/\partial l_{\mu} \neq 0$ and $\operatorname{rank}\mathcal{L}^{(0)} \equiv r$ is constant, then $\Sigma$ corresponds to a simple mode and \eqref{eikonal-eq} is its dispersion relation. On $\Sigma$, $\mathcal{L}^{(0)}$ admits $p = m - r$ linearly independent left null vector fields $\lambda_{j}$ and as many right null vector fields $\rho^{j}$, so for any integer $1 \leq j \leq p$:
\begin{equation} \label{nullvectors}
\lambda_{j}\mathcal{L}^{(0)} = 0, \quad \mathcal{L}^{(0)}\rho^{j} = 0.
\end{equation}
Correspondingly, linearly dependent null vector fields $\tilde{\lambda}_{k},\, \tilde{\rho}_{k}$ are duals of the preceding fields provided that $p + 1\leq k \leq m$. Let $(\lambda_{1}, \dots, \lambda_{p}, \tilde{\lambda}_{p+1}, \dots, \tilde{\lambda}_{m})$ and $(\rho^{1}, \dots, \rho^{p}, \tilde{\rho}^{p+1}, \dots, \tilde{\rho}^{m})$ each be a basis (in the appropriate linear space). We may then write
\begin{equation} \label{amlitude-decomposition}
A^{(n)} = a^{(n)}_{j}\rho^{j} + b^{(n)}_{j}\tilde{\rho}^{j} \equiv A^{(n)}_{1} + A^{(n)}_{2}.
\end{equation}
Based on the different parts  the two terms of $A^{(n)}$ play in the WKB expansion, we shall call $A^{(n)}_{1}$ the primary, $A^{(n)}_{2}$ the secondary amplitude of $n$th order. The zero-order equation \eqref{zeroth-order} then requires
\begin{equation}
A^{(0)} = A^{(0)}_{1} = a^{(0)}_{j}\rho^{j}
\end{equation}
thus $b^{(0)}_{j} = 0$ and $A^{(0)}$ is independent of the choice of the basis $(\rho^{j})$.

Geometrical optic can also be formulated in terms of ray tracing which is an ODE model. Provided that $l^{\mu}$ is smooth, it corresponds to locally solving the eikonal equation \eqref{eikonal-eq} through the method of bicharacteristics. Bicharacteristics of \eqref{perturbation-governing-in-Landau-gauge-eq} or so-called rays along which the amplitudes $A^{(0)}$ are transported are spacetime projections of the solutions of Hamilton's system of ODEs
\begin{equation} \label{Hamilton-equations}
\dot{x}^{\mu} = \frac{\partial\mathcal{H}}{\partial l_{\mu}}, \quad \dot{l}^{\mu} = \frac{\partial\mathcal{H}}{\partial x_{\mu}}.
\end{equation}
As long as $\psi$ is a real solution of \eqref{nullvectors}, it determines a ray bundle in \textbf{x}-space generated by the transport vector field
\begin{equation}
T^{\mu} = \frac{\partial\mathcal{H}}{\partial l_{\mu}}.
\end{equation}

\subsection{\label{sec:diffops}Explicit form of differential operators $\mathcal{L}$}
Upon inserting the principal symbol from \eqref{L-operators} into \eqref{zeroth-order}, the charasteristic equation \eqref{characteristic-eq} 
takes the form 
\begin{equation}
(g^{\mu\nu}l_{\mu}l_{\nu})^{2}[(u^{\alpha}u^{\beta} - \mathcal{Q}^{\alpha\beta})l_{\alpha}l_{\beta}](u^{\gamma}l_{\gamma})^{6} = 0.
\end{equation}
and implies three modes, consequently:
\begin{enumerate}
	\item \textit{Gravitational wave} mode, given by the Hamiltonian $\mathcal{H} = \frac{1}{2}g^{\mu\nu}l_{\mu}l_{\nu}$ belonging to high-frequency waves propagating with the speed of light on null geodesic rays (with $T^{\mu} = l^{\mu}$) without dispersion or damping.
	\item \textit{Sound wave} mode, given by the Hamiltonian $\mathcal{H} = \frac{1}{2}(u^{\alpha}u^{\beta} - \mathcal{Q}^{\alpha\beta})l_{\alpha}l_{\beta}$ propagating with the speed of sound $c_{s}$ on sound rays with $T^{\mu} = \omega(k^{\alpha}/k + u^{\alpha})$.
	\item \textit{Matter} mode, given by the Hamiltonian $\mathcal{H} = \frac{1}{2}u^{\gamma}l_{\gamma}$ that do not propagate relative to the unperturbed matter. They are generated by flow lines of background matter belonging to zero-frequency perturbations carried along 'matter rays' with $T^{\mu} = \omega(k^{\alpha}/k + u^{\alpha})$.	
\end{enumerate}
Provided that the set $(e^{\mu}_{1},e^{\mu}_{2},e^{\mu}_{3}) \subset \mathbb{A}$ is an orthonormal basis in the space orthogonal to $u^{\mu}$ and $e_{3}{}^{[\mu}k^{\nu]} = 0$, in accordance with \eqref{Landau-gauge-eq2}, the basis vectors can be explicitly expressed in the space of amplitudes $\mathbb{A}$ and it is also possible to construct a dual basis in $\tilde{\mathbb{A}}$ in pursuance of \eqref{nullvectors} for each mode:
\begin{description}
	\item[Mode I]: Gravitational wave \\
	($\mathring{l}^{2} = 0$, $\operatorname{rank}\mathcal{L}_{0} = 4$, $p = 2$):
		\begin{equation} \label{Imod-sajatvektorok}
		\begin{array}{ll}
		\lambda_{1} = e_{+}^{\mu\nu} := e_{1}^{\mu}e_{1}^{\nu} - e_{2}^{\mu}e_{2}^{\nu} & \hspace{10pt} \lambda_{2} = e_{\times}^{\mu\nu} := e_{1}^{\mu}e_{2}^{\nu} + e_{2}^{\mu}e_{1}^{\nu} \\ 
		\tilde{\lambda}_{3} = e_{1}^{\mu\nu} := 2e_{1}^{(\mu}e_{3}^{\nu)} & \hspace{10pt} \tilde{\lambda}_{4} = e_{2}^{\mu\nu} := e_{2}^{\mu}e_{3}^{\nu} + e_{3}^{\mu}e_{2}^{\nu} \\
		\tilde{\lambda}_{5} = e_{3}^{\mu\nu} := e_{3}^{\mu}e_{3}^{\nu} & \hspace{10pt} \tilde{\lambda}_{6} = e_{4}^{\mu\nu} := e_{1}^{\mu}e_{1}^{\nu} + e_{2}^{\mu}e_{2}^{\nu} \\
		\rho^{1} = e^{+}_{\mu\nu} & \hspace{10pt} \rho^{2} = e^{\times}_{\mu\nu} \\
		\tilde{\rho}^{3} = e^{1}_{\mu\nu} & \hspace{10pt} \tilde{\rho}^{4} = e^{2}_{\mu\nu} \\
		\tilde{\rho}^{5} = e^{3}_{\mu\nu} & \hspace{10pt} \tilde{\rho}^{6} = e^{4}_{\mu\nu}
		\end{array}
		\end{equation}
	\item[Mode II]: Sound wave \\
	($\mathring{\omega}^{2} = c_{*}^{2}\mathring{k}^{2}$, $\operatorname{rank}\mathcal{L}_{0} = 5$, $p = 1$):
		\begin{equation} \label{IImod-sajatvektorok}
		\begin{array}{ll}
		\lambda_{1} = (\mathring{g}^{\mu\nu} + \mathcal{Q}^{\mu\nu}) - 2e_{3}^{\mu\nu} & \hspace{10pt} \tilde{\lambda}_{2} = e_{+}^{\mu\nu} \\
		\tilde{\lambda}_{3} = e_{\times}^{\mu\nu} & \hspace{10pt} \tilde{\lambda}_{4} = e_{1}^{\mu\nu} \\
		\tilde{\lambda}_{5} = e_{2}^{\mu\nu} & \hspace{10pt} \tilde{\lambda}_{6} = e_{3}^{\mu\nu} \\
		\rho^{1} = \mathcal{Q}_{\mu\nu} + e^{3}_{\mu\nu} & \hspace{10pt} \tilde{\rho}^{2} = e^{+}_{\mu\nu} \\
		\tilde{\rho}^{3} = e^{\times}_{\mu\nu} & \hspace{10pt} \tilde{\rho}^{4} = e^{1}_{\mu\nu} \\		
		\tilde{\rho}^{5} = e^{2}_{\mu\nu} & \hspace{10pt} \tilde{\rho}^{6} = e^{3}_{\mu\nu}
		\end{array}
		\end{equation}
	\item[Mode III]: Matter \\
	($\mathring{\omega}^{2} = 0$, $\operatorname{rank}\mathcal{L}_{0} = 3$, $p = 3$)
		\begin{equation} \label{IIImod-sajatvektorok}
		\begin{array}{ll}
		\lambda_{1} = e_{1}^{\mu\nu} & \hspace{10pt} \lambda_{2} = e_{2}^{\mu\nu} \\
		\lambda_{3} = \mathcal{Q}^{\mu\nu} - 2e_{3}^{\mu\nu} & \hspace{10pt} \tilde{\lambda}_{4} = e_{+}^{\mu\nu} \\
		\tilde{\lambda}_{5} = e_{\times}^{\mu\nu} & \hspace{10pt} \tilde{\lambda}_{6} = e_{3}^{\mu\nu} \\
		\rho^{1} = e^{1}_{\mu\nu} & \hspace{10pt} \rho^{2} = e^{2}_{\mu\nu} \\
		\rho^{3} = e^{3}_{\mu\nu} & \hspace{10pt} \tilde{\rho}^{4} = e^{+}_{\mu\nu} \\		
		\tilde{\rho}^{5} = e^{\times}_{\mu\nu} & \hspace{10pt} \tilde{\rho}^{6} = e^{4}_{\mu\nu}.
		\end{array}
		\end{equation}
\end{description}
Collecting each $\rho^{j}$ one by one, the primary part of leading-order amplitude can be evaluated by having recourse to decomposition \eqref{amlitude-decomposition}. The zeroth-order amplitude shall consist of the following linear combinations of basis vectors:
\begin{equation} \label{zero-order-amplitude}
A^{(0)}_{\mu\nu} = 
\begin{cases}
a^{(0)}_{+}e^{+}_{\mu\nu} + a^{(0)}_{\times}e^{\times}_{\mu\nu} & \quad \text{(Mode I)} \\[5pt]
b^{(0)}_{1}(\mathcal{Q}_{\mu\nu} + e^{3}_{\mu\nu}) & \quad \text{(Mode II)} \\[5pt]
C{}_{(\mu}e^{3}{}_{\nu)} & \quad \text{(Mode III)}
\end{cases}
\end{equation}
where $C_{\mu} = c^{(0)}_{1}e^{1}_{\mu} + c^{(0)}_{2}e^{2}_{\mu} + c^{(0)}_{3}e^{3}_{\mu}$. Practically, the number $p$ equals the dimension of the corresponding space of polarization states. To put the prior in context, one can recognize that the first mode in \eqref{zero-order-amplitude} corresponds to an arbitrary plane-wave solution in the standard transverse traceless (TT) gauge where $(a^{(0)}_{+}, a^{(0)}_{\times})$ are the amplitudes (polarization states) of the two independent components with linear polarization, and $(e^{+}_{\mu\nu}, e^{\times}_{\mu\nu})$ are the corresponding polarization tensors. The second mode partly, and the third mode fully are in the frame orthogonal to $\mathring{u}^{\mu}$ and $\mathring{k}^{\mu}$. For the particular case of dust ($p = 0$) or in case $c_{s} = 0$, there are no sound waves and second mode degenerates into the longitudinal part of the third one. In case of stiff matter ($p = \rho$), $c_{s} = 1$, sound waves propagate with the speed of light. \cite{ehlers, ehlers2}

\subsection{\label{sec:transport-eqs}Transport equation for amplitudes}
The first-order WKB equation acquired from \eqref{zeroth-order} impose
\begin{equation} \label{first-order}
\mathcal{L}^{(0)}{}^{\alpha\beta}_{\mu\nu}A_{\alpha\beta}^{(1)} + \mathcal{L}^{(1)}{}^{\alpha\beta}_{\mu\nu}A_{\alpha\beta}^{(0)} = 0
\end{equation}
to hold for zeroth- and first-order amplitudes. Applying the expressions from \eqref{L-operators}, $\mathcal{L}^{(0)}$ is annulled upon being transvected with either one of polarization tensors given in \eqref{zero-order-amplitude}. After some manipulation, one gets the pair
\begin{equation} \label{quasi-parallel-transport}
\begin{array}{lcl}
(\nabla_{\textbf{l}} + \theta/2)a^{(0)}_{+} + \frac{1}{2}e_{+}^{\mu\nu}(a^{(0)}_{+}\nabla_{\textbf{l}}e^{+}_{\mu\nu} + a^{(0)}_{\times}\nabla_{\textbf{l}}e^{\times}_{\mu\nu}) = 0 \\[5pt]
(\nabla_{\textbf{l}} + \theta/2)a^{(0)}_{\times} + \frac{1}{2}e_{\times}^{\mu\nu}(a^{(0)}_{+}\nabla_{\textbf{l}}e^{+}_{\mu\nu} + a^{(0)}_{\times}\nabla_{\textbf{l}}e^{\times}_{\mu\nu}) = 0
\end{array}
\end{equation}
which assert that the pair of primary polarization states ($a^{(0)}_{+},\, a^{(0)}_{\times}$) are transported along rays bended by the background field. Along each null geodesic ray $x^{\mu}$ with tangent $\dot{x}^{\mu} = l^{\mu}$ given in \eqref{Hamilton-equations}, the vectors $l^{\mu},\, u^{\mu}$ span a timelike two-plane, and basis vectors span its spacelike orthogonal complement. Assuming the basis vectors $(e^{\mu}_{1},\, e^{\mu}_{2})$ are such that they are transported quasi-parallel along the rays, the transport of the primary amplitudes will be bounded by
\begin{equation} \label{quasi-parallel-transport2}
(\nabla_{\textbf{l}} + \theta/2)
\left(
\begin{array}{c}
a^{(0)}_{+} \\[5pt]
a^{(0)}_{\times}
\end{array}
\right) = 0,
\end{equation}
testifying that the change of complex vector ($a^{(0)}_{+},\, a^{(0)}_{\times}$) along the ray consists solely of a rescaling as in the case of GWs in vacuum.

Making use of the expression for $\mathcal{L}^{(2)}$ from \eqref{L-operators}, the transport equation for the first-order primary amplitudes yields
\begin{equation} \label{first-order-quasi-parallel-transport}
\begin{array}{ll}
(\nabla_{\textbf{l}} + \theta/2)a^{(1)}_{+} \! = & \! \!\displaystyle \frac{1}{2}\kappa(\rho - 5p)a^{(0)}_{+} + \frac{1}{2}e_{+}^{\alpha\beta}\big[2(\nabla^{\mu}\nabla_{\alpha} + \phantom{)} \\
\mkern-9mu \phantom{(} \nabla_{\alpha}\nabla^{\mu})\delta^{\nu}_{\beta} & \mkern-63mu - \delta^{\mu}_{\alpha}\delta^{\nu}_{\beta}\nabla^{2} - c_{*}^{-2}\mathcal{P}^{\mu\nu}\nabla_{\alpha}\nabla_{\beta}\big][a^{(0)}_{+}e^{+}_{\mu\nu} + a^{(0)}_{\times}e^{\times}_{\mu\nu}]
\end{array}
\end{equation}
and a similar one for $a^{(1)}_{\times}$.

\section{\label{sec:num-sol-and-applic}Numerical solution and application for data analysis}
\subsection{\label{sec:num-solution}Numerical solution for the transport equation}
\begin{figure}[h]
	\includegraphics[scale=0.55]{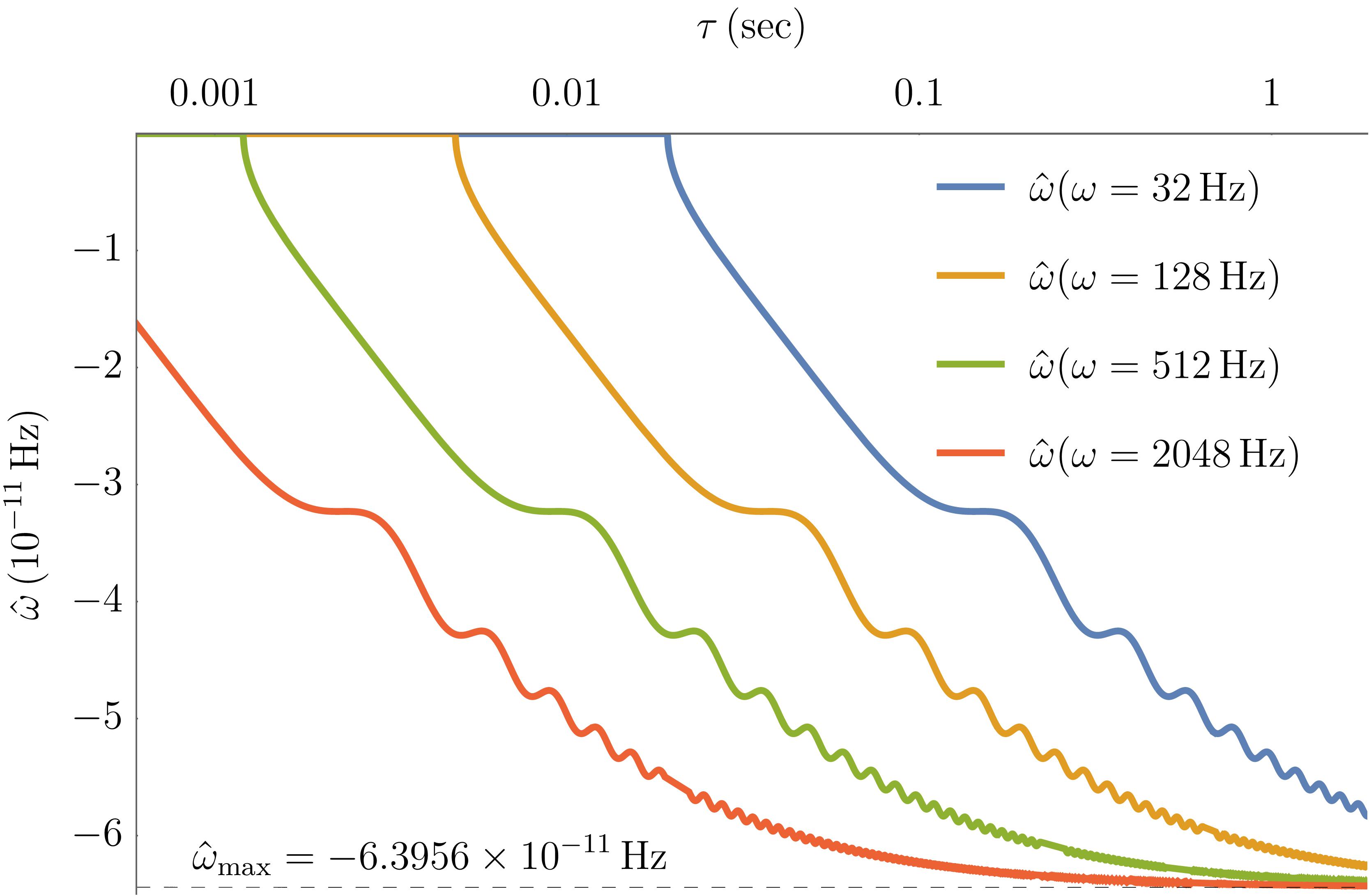}
	\caption{\label{fig:omega1} Retarded-time dependence of frequency-shift $\hat{\omega}$ for plain waves of several discrete values of initial frequency. Higher initial frequencies approach the limiting $\hat{\omega}_{\text{max}}$ exponentially faster.}
\end{figure}

Assuming the real part of the zeroth-order amplitude of the plane wave \eqref{WKB-eqs1} travelling in the equatorial plain $\vartheta = \pi/2$ in $z$-direction to be
\begin{equation} \label{planewave}
A^{(0)}_{\mu\nu} = (a^{(0)}_{+}e^{+}_{\mu\nu} + a^{(0)}_{\times}e^{\times}_{\mu\nu})\sin[\tau\omega],
\end{equation}
with the metric potentials $e^{\nu},\, e^{\lambda}$ in \cite{barta} and the retarded time
\begin{equation}
\tau = -t e^{\nu(z)} + ze^{\lambda(z)},
\end{equation}
the transport equation \eqref{first-order-quasi-parallel-transport} reduces to the form of
\begin{equation} \label{reduced-form}
	\begin{array}{l}
	    \displaystyle \left([64C_{2}\hat{\omega}^2 + C_{1}\omega - 2\lambda']a^{(0)}_{+} + C_{0}\hat{\omega}a^{(0)}_{\times}\right)\cos\left(\tau\hat{\omega}\right) \\[8pt]
	    \displaystyle + \left(4[S_{2}\omega + S_{1}]\hat{\omega}a^{(0)}_{+} +  3\lambda'^2 a^{(0)}_{\times}\right)\sin\left(\tau\hat{\omega}\right) \\[8pt]
	    = \displaystyle 2\kappa(\rho - 5p)e^{3\lambda}a^{(0)}_{+}
	\end{array}
\end{equation}
where primes denote partial differentiation with respect to coordinate $z$ and
\begin{equation}
	\begin{array}{ll}
		C_{2} = & \mkern-9mu e^{2\lambda}(1+z\lambda')^2 + t^2e^{2\nu}\nu'^2 - e^{\lambda+\nu}\left(1+2\nu't\left[1+z\lambda'\right]\right) \\
		
		C_{1} = &\mkern-9mu e^{4\lambda}[\nu' + \lambda'(19+11z\lambda'+z\nu') + 8z\lambda''] \\
		
		C_{0} = &\mkern-9mu 24\lambda'\left[e^{\lambda}(1+z\lambda') - t\nu'e^{\nu}\right] \\
		&\mkern-9mu - 24te^{3\lambda+\nu}(3\nu'\lambda' + 9\nu'^2 + 8\nu'') \\
		
		S_{2} = &\mkern-9mu e^{5\lambda}(1+z\lambda')^2 - 2te^{4\lambda+\nu}(1+z\lambda')\nu' + e^{3\lambda+2\nu}(1+t^2\nu'^2) \\
		
		S_{1} = &\mkern-9mu 16e^{\lambda}(2\lambda+z\lambda'^2+z\lambda'') - 16te^{\nu}(\nu'^2-\nu'')
	\end{array}
\end{equation}
are functions of $(t,\,z)$ alone. For the sake of the simpler representation of the following results, the coordinate time $t$ is to be replaced by the retarded time $\tau$. The numerical solution for $\hat{\omega}$ given by \eqref{reduced-form} for any possible values of ($\tau,\,\omega$) is expected to be negative, owing to the energy dissipation, and to be extremely small compared to the initial frequency $\omega$. Having the eqs. (\ref{line-element}--\ref{EoS}) applied for a nebula of radius $R = 50$ pc \cite{murray,barta}, the maximal decrease in frequency is found to be $\hat{\omega}_{\text{max}} = -6.3956\times 10^{-11} $ Hz. Here and thereafter the order of magnitude of amplitudes $(a^{(0)}_{+},\, a^{(0)}_{\times})$ is set to $\mathcal{O}(10^{-21})$. All the frequency components that make up the waveform suffer a tiny, gradually decreasing 'redshift-like' alteration $\hat{\omega}$ that depends on the initial frequency and the position of wavefront. 
Fig.~\ref{fig:omega1} displays the frequency-shift $\hat{\omega}$ on a log scale for several discrete values of initial frequency for an arbitrary plane wave \eqref{planewave} that lasts for $\tau = 2$ sec. The figure reveals a characteristic distinctive feature of self-similarity of the frequency-shift curves for different initial frequencies.

\begin{figure}[h]
	\includegraphics[scale=0.70]{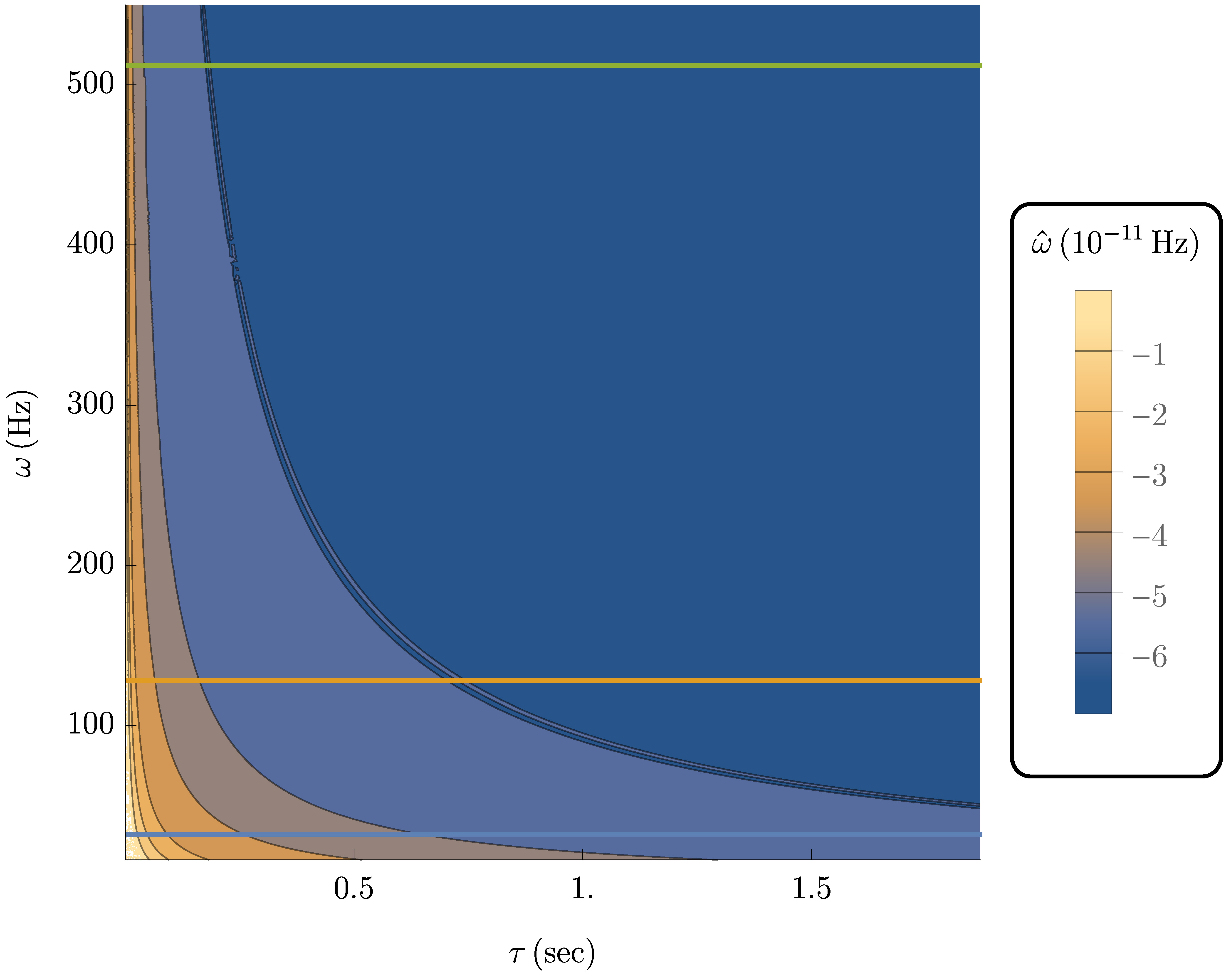}
	\caption{\label{fig:omega2} Frequency shift in the low-frequency regime. The horizontal lines represent the first three initial frequency values ($32$ Hz, $128$ Hz, $512$ Hz) shown in Fig.~\ref{fig:omega1} with respective colours.}
\end{figure}

A more detailed figure, Fig.~\ref{fig:omega2} demonstrates the frequency shifts that belong to any initial frequency in the $(\tau,\, \omega)$ domain. 
For the gravitational-wave transient GW150914, the effect is the greatest in the frequency range $100 \text{--} 250$ Hz.

\subsection{\label{sec:gw-analysis}Matched filtering techniques for gravitational-wave data analysis}
Matched filtering is a data analysis technique that allows us to efficiently extract faint gravitational-wave signals of known form from a noise-dominated data. The technique is based on correlating the output of detectors with waveform templates. In a similar way, we will investigate the correlation of a theoretically given template waveform $(h_{1})$ with the one that have suffered alteration $(h_{2})$ under the influence of the medium. As the Advanced LIGO and Advanced Virgo detectors can follow the phase $\phi$ of a gravitational-wave signal in time, the time series (or time-domain waveforms) $h(t)$ is often represented in the frequency domain in terms of the power-spectral density (PSD) $S_{n}(f)$ due to instrument noise (see in \cite{LIGO2}) and the Fourier transform
\begin{equation}
\tilde{h}^{*}(f) = \int_{-\infty}^{+\infty}\mathrm{e}^{\mathrm{-2\pi i} f t}h(t)\,\mathrm{d}t
\end{equation}
of the time series $h(t)$. We define the noise-weighted inner product of two time-domain waveforms $h_{1}(t)$ and $h_{2}(t)$ as
\begin{equation} \label{inner-product}
	\langle h_{1}|h_{2} \rangle = 4\Re\int\limits_{f_{\mathrm{min}}}^{f_{\mathrm{max}}}\frac{\tilde{h}^{*}_{1}(f)\tilde{h}_{2}(f)}{S_{n}(f)}\,\mathrm{d}f
\end{equation}
where the limits of integration ($f_{\mathrm{min}},\, f_{\mathrm{max}}$) correspond to the upper and lower sidebands of the detector. The expectation value of the optimal matched filtering signal-to-noise ratio (SNR), given as
\begin{equation} \label{SNR}
\frac{S}{N} = \frac{\langle h_{1}|h_{2} \rangle}{\sqrt{\langle h_{1}|h_{1} \rangle}},
\end{equation}
is as high as ca. $60$ gets more likely to be detected. Another useful mean for examining differences in waveforms is to measure their overlap $\mathcal{O}$, which is
\begin{equation} \label{overlap}
\mathcal{O} = \max_{t_{0},\psi_{0}}\frac{\langle h_{1}|h_{2} \rangle}{\sqrt{\langle h_{0}|h_{0} \rangle\langle h_{1}|h_{1} \rangle}} \equiv 1 - \mathcal{M}.
\end{equation}
The overlap is a normalized SNR maximalized over the initial time $t_{0}$ and phase $\psi_{0}$ of the template waveform. It is also related to the mismatch $\mathcal{M}$ between signal and template. For a straight forward derivation of eqs. (\ref{inner-product}--\ref{overlap}) see \cite{apostolatos}.

\begin{figure}[h]
	\includegraphics[scale=0.70]{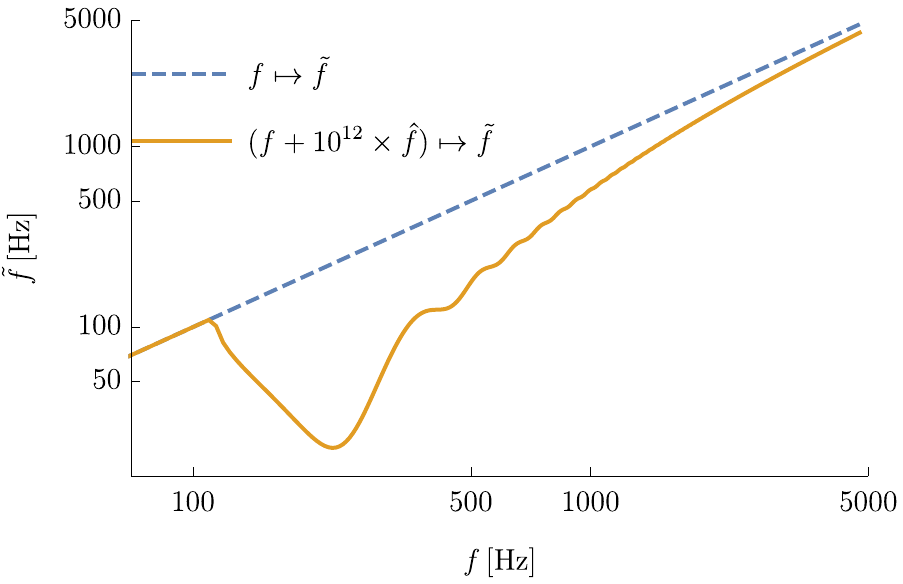}
	\caption{\label{fig:omega3} 
		The dashed blue line represents the identity map $f \mapsto f$ whereas the solid orange line shows the expected deviation in the frequency for transient event GW150914 after it crossed the medium. To illustrate the nature of the frequency-shift, $\hat{f}$ was amplified by a factor of $10^{12}$.}
\end{figure}

In this paper, the power-spectral density $S_{n}(f)$ of the detector noise was taken from the average-measured strain-equivalent noise, or sensitivity, of the Advanced LIGO detectors at \href{https://losc.ligo.org/s/events/GW150914/P1500238/H1-GDS-CALIB_STRAIN.txt}{Hanford (H1)} and \href{https://losc.ligo.org/s/events/GW150914/P1500238/L1-GDS-CALIB_STRAIN.txt}{Livingston (L1)} sites (within bandwidth $0.125 \text{--} 8192$ Hz) at the time the gravitational-wave event designated as 'GW150914' was observed. The template $h_{1}(f)$ represents the \href{https://losc.ligo.org/s/events/GW150914/P150914/fig2-unfiltered-template-reconstruction-H.txt}{reconstructed time series} of the gravitational-wave transient signal that was released for event GW150914 by the LIGO Scientific Collaboration and Virgo Collaboration \cite{LIGO1,LIGO2}. While $h_{2}(\tilde{f}) \equiv h_{1}(f + \hat{f})$ stands for the altered signal where $\hat{f} \equiv \hat{\omega}/2\pi$. The change of frequency in the entire frequency band, specific for transient event GW150914 is shown in Fig.~\ref{fig:omega3}. The deviation of $\tilde{f}$ from $f$ is most prominently observed in the interval of $110 \text{--} 380$ Hz in the frequency domain, designated as super-low frequency (SLF). The rapid growth of $\tilde{f}$ starting at $f=110$ Hz comes to a standstill at a major low peak at $f=220$ Hz. From that point on, $\tilde{f}$ approaches $\hat{\omega}_{\text{max}}$ at a much slower rate. It also comes to one's attention that in the kHz regime, the deviation is nearly comparable with the original frequency $f$.

\begin{figure}[t]
	\includegraphics[scale=0.70]{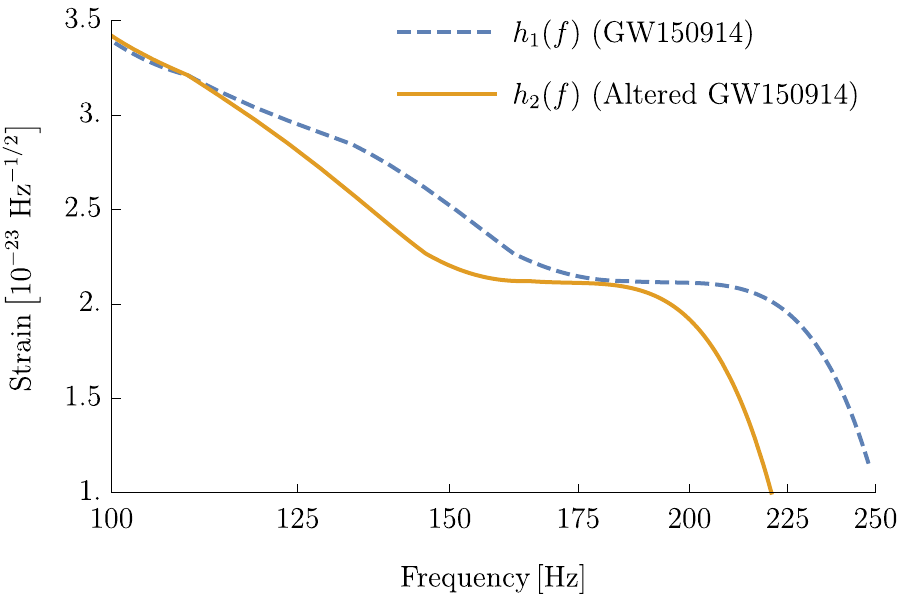}
	\caption{\label{fig:omega4} The upper peak envelope (UPE) of the coincident signal in GW150914 $h_{1}(f)$ (dashed blue line) and its frequency-altered counterpart $h_{2}(f)$ (solid orange line) in frequency regime above 100 Hz. Both are fitted on maxima points by cubic spline interpolation.}
\end{figure}

Due to the non-linear change in frequency, the signal distortion exhibits a complex behaviour. Fig.~\ref{fig:omega4} shows the magnitude of the original and the altered signal from GW150914 ($h_{1}$ and $h_{2}$, respectively) versus the frequency in the band $100 \text{--} 250$ Hz.   

\begin{figure}[t]
	\includegraphics[scale=0.70]{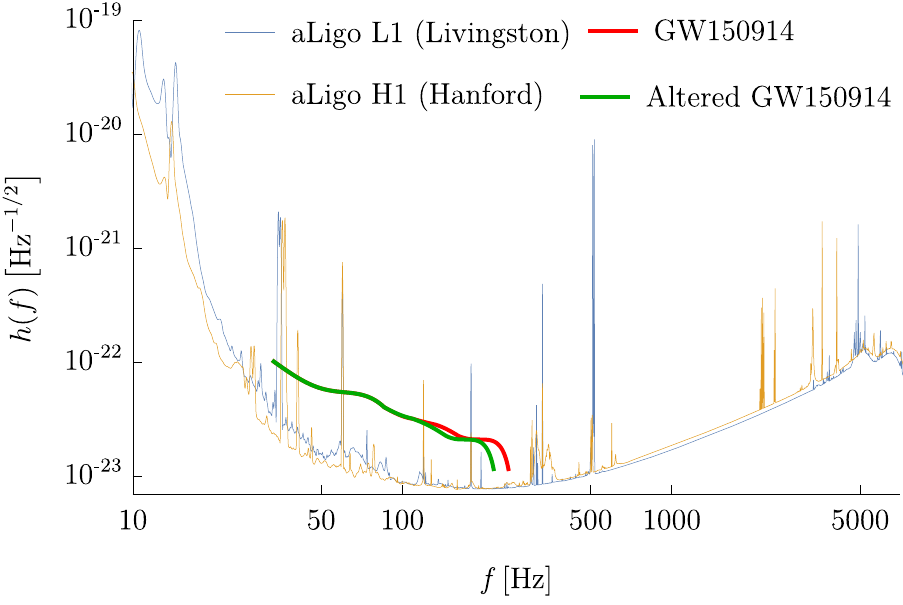}
	\caption{\label{fig:omega5} Gravitational-wave strain amplitude from GW150914 projected onto sensitivity curve of $H1$ (dashed blue line) and frequency-altered counterpart (solid red line) in the full bandwidth of the detector.}
\end{figure}

To measure the actual difference between the original signal and the frequency-altered counterpart we calculated the overlap \eqref{overlap} for Hanford detector's PSD (due to its better sensitivity in low-frequencies compared to $L1$). Consequently, the overlap is $\mathcal{O}_{H1} = 0.970995$ for a signal $h_{2}(f)$ in which the contribution of $\hat{f}$ was amplified by a factor of $10^{12}$. Apart from demonstrational purposes (cf. Figs.~\ref{fig:omega3}, \ref{fig:omega4}) the amplification was required to appropriately increase the numeric working precision which would not have been large enough to suppress numerical errors otherwise. Fig.~\ref{fig:omega5} shows the UPEs of $h_{1}(f)$ and $h_{2}(f)$ (same as in Fig.~\ref{fig:omega4}) projected onto sensitivity curves of aLIGO detectors.

\section{Conclusions and summary of results}
In order to provide a more accurate picture of expected waveforms for direct detection, we have carried out a general study on the interaction of gravitational waves with matter. We have considered the wave passing through a vast spherical assemblage of cold, compressible gas, called a giant molecular cloud. Gravitational waves were treated as linearized metric perturbations embedded in an interior Schwarzschild spacetime that belongs to nebula. The perturbed quantities lead to the field equations governing the gas dynamics and describe the interaction of gravitational waves with matter. The field equations decoupled to a set of PDEs of different orders of magnitude by WKB approximation, assuming the GW-amplitude to relatively slowly vary compared to the rapid oscillation of the phase. In the frame of WKB approximation, the dispersion relation indicates three distinct degenerate modes of polarization. Two of them are regular and correspond to gravitational and sound waves obeying the transport equations along rays (determined via Hamilton's ODEs on characteristic hypersurfaces), whereas the zero-frequency one is singular and represents non-propagating density and vorticity perturbations of the dispersive medium. See corresponding result in \cite{ehlers,ehlers2,prasanna}. In regular case, the primary amplitudes follow null-characteristics (see Sec. \ref{sec:eikonal}), whereas the obtained transport equation of secondary amplitudes depends upon the material density in geometric optics limit.

The principal result established in this paper is the demonstration that in the framework of post-geometrical optics, the transport equation of secondary amplitudes provides numerical solutions for the frequency-alteration $\hat{f}$. On the grounds that the energy dissipating process is responsible for decreasing frequency, $\hat{f}$ is bound to be negative, yet extremely small compared to the unaltered frequency $f$.

As an illustrative example, we considered a nebula with a mass assigned to an average-sized giant molecular cloud (GMC), namely $10^5$ $M_{\odot}$ and diameter of $100$ parsecs (typically ranges $5 \text{--} 200$ parsecs). Whereas the average density in the solar vicinity is one particle per cubic centimeter, the average density of a GMC is a hundred to a thousand times as great. (See, e.g. Table 1 and the Appendix of \cite{murray}.) In fact, even in such a dense environment, for any unaltered frequency that falls within the bandwidth of current advanced ground-based detectors the frequency alteration still remains so small (circa $10^{-11}$ Hz) that its influence is practically untraceable: the resulting mismatch $\mathcal{M}$ between measurement and expectation was barely $2.9005 \times 10^{-14}$. However, the frequency-alteration pattern exhibits a power-law relationship between $f$ and $\hat{f}$ which implies that, with further increasing sensitivity and bandwidth, the analysis of higher frequency signals in future may not disregard this contribution. 

For sources in the $1\text{--}2$ kHz frequency range, the influence of the interaction on the signal may increase by as high as 6 orders of magnitude compared to that of the value on initial frequency of $100 \text{--} 200$ Hz. Such high-frequency signals are expected to be emitted from the post-merger phase of low-mass binary-neutron-star (BNS) coalescence.  The dominant frequency of the post-merger signal of two $1.2$ $M_{\odot}$ mass NSs (with the LS220 equation of state) was computed at $2.56$ kHz by relativistic simulations. \cite{foucart} It is also noteworthy that mergers of the above-mentioned sources are detectable over much greater distances than those of stellar-mass BBHs such as GW150914. Moreover, even higher frequencies ($3 \text{--} 4$ kHz) of quasi-periodic signals are expected from the formation of the hypermassive neutron stars. \cite{shibata} 

Another possibility to significantly 'boost' the effect is to increase the density of the environment by taking active galactic nuclei (AGN) into consideration. 
Suppose that an event of dynamical merger occurs in the central region of an AGN and it is observed over its dissipative accretion disk. In this extremely favorable 
but rare angular position, $\hat{f}$ might be sufficiently large to be measured.

\begin{acknowledgments}
This research has made use of data, software and/or web tools obtained from the LIGO Open Science Center \href{https://losc.ligo.org}{(https://losc.ligo.org)}, a service of LIGO Laboratory and the LIGO Scientific Collaboration. LIGO is funded by the U.S. National Science Foundation. M. V. was supported by the J\'{a}nos Bolyai Research Scholarship of the Hungarian Academy of Sciences. Partial support comes from “NewCompStar,” COST Action Program MP1304.
\end{acknowledgments}

\bibliography{cikk2}

\end{document}